\documentstyle[12pt,epsf]{article}

\renewcommand{\bar}[1]{\overline{#1}}

\newcommand{\etal}{{\em et al.}}
\newcommand{\ie}{{\em i.e.}}

\def\overdoublearrow#1{\vbox{\ialign{##\crcr
    $\leftrightarrow$\crcr\noalign{\kern0pt\nointerlineskip}
    $\hfil\displaystyle{#1}\hfil$\crcr}}}

\begin{document}

\begin{flushright}
{\small
SLAC--PUB--8663\\
October 2000\\}
\end{flushright}

\vfill

\begin{center}
{{\bf\LARGE   
Two-Photon Exclusive Processes in QCD}\footnote{Work
supported by the Department of Energy
under contract number DE-AC03-76SF00515.}}

\bigskip
Stanley J. Brodsky\\
{\sl Stanford Linear Accelerator Center \\
Stanford University, Stanford, California 94309\\
sjbth@slac.stanford.edu}\\
\medskip
\end{center}

\vfill

\begin{center}
{\bf\large   
Abstract }
\end{center}

Exclusive two-photon reactions such as Compton scattering at large
angles, deeply virtual Compton scattering, and hadron production in
photon-photon collisions provide important tests of QCD at the
amplitude level, particularly as measures of hadron distribution
amplitudes and skewed parton distributions.

\vfill

\begin{center} 
{\it Invited talk presented at Photon 2000} \\
{\it Ambleside, England }\\
{\it August 26--31, 2000}\\
\end{center}

\vfill

\newpage


\renewcommand{\baselinestretch}{1.2}
\normalsize

\section{Introduction}

A central focus of study in QCD are the wavefunctions which
describe hadrons in terms of their quark and gluon degrees of freedom at
the amplitude level.  Of particular interest are the gauge- and
process-independent meson and baryon valence-quark distribution amplitudes
$\phi_M(x,Q)$, and
$\phi_B(x_i,Q)$ which control exclusive processes involving a hard scale
$Q$; for example, meson
distribution amplitudes play a key role in the analysis of exclusive
semi-leptonic and two-body hadronic
$B$-decays\cite{BHS,Sz,BABR,Beneke:1999br,Keum:2000ph,Keum:2000wi}.
There has recently been considerable progress both in calculating
hadron wavefunctions from first principles in QCD and in measuring them
using diffractive di-jet dissociation.

Two-photon processes such as $\gamma^* \gamma
\to $ hadrons, Compton scattering $\gamma p \to \gamma p$ at large
momentum transfer, and
$\gamma
\gamma
\to$ hadron pairs at high momentum transfer and fixed
$\theta_{cm}$, can play a crucial
role in understanding the perturbative and non-perturbative structure of
QCD, first by testing the validity and empirical applicability of
leading-twist factorization theorems, second by verifying the structure
of the underlying perturbative QCD subprocesses,
and third, through
measurements of angular distributions and ratios which are
sensitive to the shape of the distribution amplitudes.
In effect, Compton scattering and photon-photon collisions
are microscopes for testing fundamental
scaling laws of PQCD and for measuring distribution amplitudes.  In
addition, as I shall discuss in the next section, deeply virtual Compton
scattering
$\gamma^* p
\to
\gamma p$ for far off-shell initial photons has
emerged as one of the most important and interesting exclusive QCD
reactions.

\section{Deeply Virtual Compton Scattering}

The virtual Compton scattering amplitude ${d\sigma\over dt}(\gamma^*
p \to \gamma p)$ has extraordinary sensitivity to
fundamental features of proton
structure\cite{Ji:inclusive,Radyushkin:1997ki,%
Guichon:1998xv,Vanderhaeghen:1998uc,%
Collins:1999be,Diehl:combined,%
Blumlein:2000cx,Penttinen:2000dg}.
Even though the final state photon is on-shell, the deeply virtual
Compton process probes the elementary quark structure of the proton near
the light cone as an effective local current.  In contrast to deep
inelastic scattering, which measures only the absorptive part of the
$t = 0$ forward virtual Compton amplitude, deeply virtual Compton
scattering allows the measurement of the phase and spin structure of
proton matrix elements for general momentum transfer $t$.  The scaling,
Regge behavior, and phase structure of deeply virtual Compton scattering
have been discussed in the context of the covariant parton model in Ref.
\cite{BCG7273}.  The interference of Compton and
bremsstrahlung amplitudes gives an electron-positron asymmetry in
the ${e^\pm  p \to e^\pm \gamma p}$ cross section which is proportional
to the real part of the Compton amplitude\cite{BCG7273}.

To leading order in $1/Q$, the deeply virtual Compton scattering
amplitude factorizes as the convolution in $x$ of the
amplitude $t^{\mu \nu}$ for hard Compton scattering on a quark line with
the generalized Compton form factors $H(x,t,\zeta),$ $ E(x,t,\zeta)$,
$\tilde H(x,t,\zeta),$ and $\tilde E(x,t,\zeta)$ of the target
proton.
Here
$x$ is the light-cone momentum fraction of the struck quark, and
$\zeta= Q^2/2 P\cdot q$ plays the role of the Bjorken variable.
The form factor $H(x,t,\zeta)$ describes the proton response
when the helicity of the proton is unchanged, and
$E(x,t,\zeta)$ is for the case when the proton helicity is flipped.  Two
additional functions $\tilde H(x,t,\zeta),$ and $\tilde E(x,t,\zeta)$
appear, corresponding to the dependence of the Compton amplitude on
quark helicity.  These ``skewed" parton distributions involve non-zero
momentum transfer, so that a probabalistic interpretation
is not possible.  However, there are remarkable sum
rules connecting the chiral-conserving and chiral-flip form factors
$H(x,t,\zeta)$ and $ E(x,t,\zeta)$ with the corresponding spin-conserving
and spin-flip electromagnetic form factors $F_1(t)$ and $F_2(t)$ and
gravitational form factors $A_{\rm q}(t)$ and $B_{\rm q}(t)$ for each
quark and anti-quark constituent\cite{Ji:inclusive}.
Thus deeply virtual Compton scattering is related to the quark
contribution to the form factors of a proton scattering in a
gravitational field.

One can construct space-like electromagnetic,
electroweak, gravitational couplings,  or any local operator product matrix
element from the diagonal overlap of the LC wavefunctions
\cite{Brodsky:1980zm}.
In the case of the generalized form factors of deeply
virtual Compton scattering, the
computation\cite{Brodsky:2000xy,Diehl:2000xz} requires not only the
diagonal matrix element
$n \rightarrow n$ for $\zeta < x <1$, where parton number is conserved,  but
also an off-diagonal
$n+1\rightarrow n-1$ convolution for $0< x < \zeta$.  This second domain
occurs since the current operator of the final-state photon with
positive light-cone momentum fraction
$\zeta$ can annihilate a
$q{\bar{q'}}$ pair in the initial proton wavefunction.  The off-diagonal
terms are referred to in the literature as the ``ERBL" contributions,
since they resemble virtual Compton scattering on an exchanged mesonic
system $\gamma^* q{\bar{q'}}\to \gamma$ and thus obey the same evolution
equations in $\log q^2$ as the meson distribution
amplitudes
\cite{Lepage:1979zb,Brodsky:1979qm,Lepage:1980fj,Efremov:1980rn}.  In
fact, the light cone Fock representation shows that there are
underlying relations between the Fock states of different particle number
which interrelate the two domains.

\section{Non-Perturbative Calculations of the Pion Distribution Amplitude}

The distribution amplitude $\phi(x,\widetilde Q)$ can be computed from
the integral over transverse momenta of the renormalized hadron valence
wavefunction in the light-cone gauge at fixed light-cone time
\cite{BrodskyLepage}:
\begin{equation}
\phi(x,\widetilde Q) = \int d^2\vec{k_\perp}\thinspace
\theta \left({\widetilde Q}^2 - {\vec{k_\perp}^2\over x(1-x)}\right)
\psi^{(\widetilde Q)}(x,\vec{k_\perp}),
\label{quarkdistamp}
\end{equation}
where a global cutoff in invariant mass is identified with the resolution
$\tilde Q$.  The distribution amplitude $\phi(x, \tilde Q)$ is boost and
gauge invariant and evolves in $\ln \tilde Q$ through an evolution
equation\cite{Lepage:1979za,Lepage:1979zb,Lepage:1980fj}.
Since it is formed
from the same product of operators as the non-singlet structure function,
the anomalous dimensions controlling $\phi(x,Q)$ dependence in the
ultraviolet
$\log Q$ scale are the same as those which appear in the DGLAP
evolution of structure functions\cite{Brodsky:1980ny}.
The decay $\pi \to \mu \nu$ normalizes the wave function at the origin:
${a_0/ 6} = \int^1_0 dx \phi(x,Q) = {f_\pi/ (2 \sqrt 3)}.$
One can also compute the distribution amplitude from the gauge invariant
Bethe-Salpeter wavefunction at equal light-cone time.  This also allows
contact with both QCD sum rules\cite{Shifman:1979by} and lattice
gauge theory; for example, moments of the pion distribution amplitudes
have been computed in lattice gauge theory
\cite{Martinelli:1987si,Daniel:1991ah,DelDebbio:2000mq}.
Conformal symmetry can be used as a template to organize the
renormalization scales and evolution of QCD
predictions \cite{Brodsky:1980ny,Brodsky:2000cr}.
For example,   Braun and collaborators have shown how one can use
conformal symmetry to classify the eigensolutions of the baryon
distribution amplitude\cite{Braun:1999te}.

\begin{figure}[htbp]
\begin{center}
\leavevmode
{\epsfxsize=5.5in\epsfbox{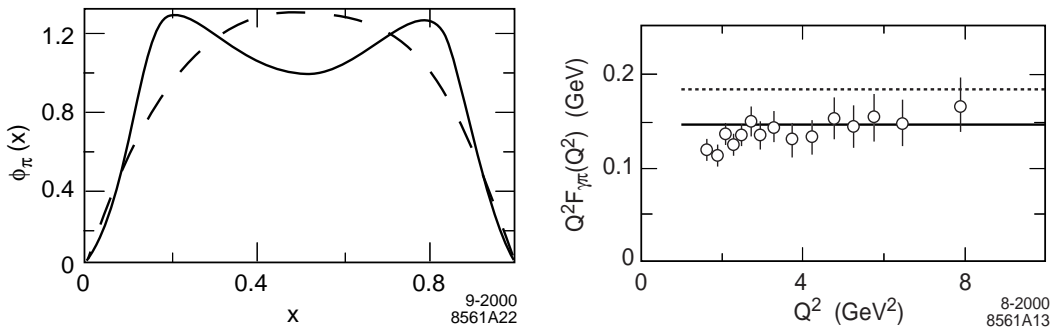}}
\end{center}
\caption[*]{
(a) Preliminary transverse lattice results for the pion distribution
amplitude at $Q^2
\sim 10 {\rm GeV}^2$.  The solid curve is the theoretical
prediction from the combined DLCQ/transverse lattice
method\cite{Dalley:2000dh}; the chain line is the experimental result
obtained from jet diffractive dissociation\cite{Ashery:1999nq}.  Both are
normalized to the same area for comparison.
(b) Scaling of the transition photon to pion transition form
factor $Q^2F_{\gamma \pi^0}(Q^2)$.  The dotted and
solid theoretical curves are the perturbative QCD prediction at leading
and next-to-leading order, respectively, assuming the asymptotic pion
distribution The data are from the CLEO
collaboration\cite{Gronberg:1998fj}.}
\label{Fig:DalleyCleo}
\end{figure}

Dalley\cite{Dalley:2000dh} has recently calculated the pion
distribution amplitude from QCD using a combination of the discretized
light-cone quantization\cite{dlcq} method for the $x^-$
and
$x^+$ light-cone coordinates with the transverse
lattice method\,\cite{bard,mat2} in the transverse directions,  A finite
lattice spacing $a$ can be used by choosing
the parameters of the effective theory in a region of
renormalization group stability to respect the required gauge,
Poincar\'e, chiral, and continuum symmetries.
The overall normalization gives $f_{\pi} = 101$ MeV
compared with the experimental value of $93$ MeV.
Figure \ref{Fig:DalleyCleo} (a)
compares the resulting DLCQ/transverse lattice pion wavefunction with
the best fit to the diffractive di-jet data (see the next section) after
corrections for hadronization and experimental acceptance
\cite{Ashery:1999nq}.  The theoretical curve
is somewhat broader than the experimental result.  However, there
are experimental uncertainties from hadronization and
theoretical errors introduced from finite DLCQ resolution,
using a nearly massless pion, ambiguities in setting the factorization
scale
$Q^2$, as well as errors in the evolution of the distribution amplitude
from 1 to $10~{\rm GeV}^2$.  Instanton models also predict a pion
distribution
amplitude close to the asymptotic form\cite{Petrov:1999kg}.
In contrast,  recent lattice results from Del Debbio
{\em et al.}\cite{DelDebbio:2000mq} predict a much narrower
shape for the pion distribution amplitude than the distribution predicted
by the transverse lattice.
A new result for the proton distribution amplitude treating nucleons as
chiral solitons
has recently been derived by Diakonov and Petrov\cite{Diakonov:2000pa}.
Dyson-Schwinger models\cite{Hecht:2000xa} of
hadronic Bethe-Salpeter wavefunctions can also be used to
predict light-cone wavefunctions and hadron distribution amplitudes by
integrating over the relative $k^-$ momentum.  There is also the
possibility of deriving Bethe-Salpeter wavefunctions within light-cone
gauge quantized QCD\cite{Srivastava:2000gi} in order to properly match to
the light-cone gauge Fock state decomposition.

\vspace{.5cm}

\section{Measurements of the Pion Distribution Amplitude by Di-jet
Diffractive Dissociation} The shape of hadron distribution
amplitudes can be measured in the diffractive
dissociation of high energy hadrons into jets on a nucleus.
For example, consider the
reaction\cite{Bertsch,MillerFrankfurtStrikman,Frankfurt:1999tq}
$\pi A \rightarrow {\rm Jet}_1 + {\rm Jet}_2 + A^\prime$
at high energy where the nucleus $A^\prime$ is left intact in its ground
state.  The transverse momenta of the jets balance so that
$
\vec k_{\perp i} + \vec k_{\perp 2} = \vec q_\perp < {R^{-1}}_A \ .
$
The light-cone longitudinal momentum fractions also need to add to
$x_1+x_2 \sim 1$ so that $\Delta p_L < R^{-1}_A$.  The process can
then occur coherently in the nucleus.  Because of color transparency and
the long coherence length,  a valence $q \bar q$ fluctuation of the pion
with small impact separation will penetrate the nucleus with minimal
interactions, diffracting into jet pairs\cite{Bertsch}.  The
$x_1=x$,
$x_2=1-x$ dependence of the di-jet distributions will thus reflect the
shape of the pion valence light-cone wavefunction in $x$; similarly, the
$\vec k_{\perp 1}- \vec k_{\perp 2}$ relative transverse momenta of the
jets gives key information on the second derivative of the underlying
shape of the valence pion
wavefunction\cite{MillerFrankfurtStrikman,Frankfurt:1999tq,BHDP}.
The diffractive nuclear amplitude extrapolated to
$t = 0$ should be linear in nuclear number $A$ if color transparency is
correct.  The integrated diffractive rate should then scale as $A^2/R^2_A
\sim A^{4/3}$.

The E791 collaboration at Fermilab has recently
measured the diffractive di-jet dissociation of 500 GeV incident pions
on nuclear targets\cite{Ashery:1999nq}.  The results are consistent with
color transparency, and the momentum partition of the
jets conforms closely with the shape of the asymptotic distribution
amplitude,
$\phi^{\rm asympt}_\pi (x) =
\sqrt 3 f_\pi x(1-x)$, corresponding to the leading anomalous dimension
solution\cite{Lepage:1979zb,Lepage:1980fj} to the perturbative QCD
evolution equation.

\section{The Photon-to-Pion Transition Form Factor and the Pion
Distribution Amplitude}

The simplest and perhaps most elegant illustration of an exclusive
reaction in QCD is the evaluation of the photon-to-pion transition form
factor $F_{\gamma \to \pi}(Q^2)$ which is measurable in single-tagged
two-photon
$ee \to ee \pi^0$ reactions.
The form factor is
defined via the invariant amplitude
$
\Gamma^\mu = -ie^2 F_{\pi \gamma}(Q^2) \epsilon^{\mu \nu \rho \sigma}
p^\pi_\nu \epsilon_\rho q_\sigma \ .$
As in inclusive
reactions, one must specify a factorization scheme which divides the
integration regions of the loop integrals into hard and soft momenta,
compared to the resolution scale $\tilde Q$.
At leading twist, the transition form factor then factorizes as a
convolution of the
$\gamma^* \gamma \to q \bar q$ amplitude (where the quarks are
collinear with the final state pion) with the valence light-cone
wavefunction of the pion:
\begin{equation}
F_{\gamma M}(Q^2)= {4 \over \sqrt 3}\int^1_0 dx \phi_M(x,\tilde Q)
T^H_{\gamma \to M}(x,Q^2) .
\label{transitionformfactor}
\end{equation}
The hard scattering amplitude for $\gamma\gamma^*\to q \bar q$
is
$
T^H_{\gamma M}(x,Q^2) = { [(1-x) Q^2]^{-1}}\left(1 +
{\cal O}(\alpha_s)\right).
$
The leading QCD corrections have been computed by Braaten
\cite{Braaten}.
The evaluation of the next-to-leading corrections in the physical
$\alpha_V$ scheme is given in Ref. \cite{Brodsky:1998dh}.
For the
asymptotic distribution amplitude $\phi^{\rm asympt}_\pi (x) =
\sqrt 3 f_\pi x(1-x)$ one predicts
$
Q^2 F_{\gamma \pi}(Q^2)= 2 f_\pi \left(1 - {5\over3}
{\alpha_V(Q^*)\over \pi}\right)$ where $Q^*= e^{-3/2} Q$ is the BLM scale
for the pion form factor.  The PQCD predictions have
been tested in measurements of $e \gamma \to e \pi^0$ by the CLEO
collaboration\cite{Gronberg:1998fj}.  See Fig.
\ref{Fig:DalleyCleo} (b).
The
flat scaling of the $Q^2 F_{\gamma \pi}(Q^2)$ data from $Q^2 = 2$
to $Q^2 = 8$ GeV$^2$ provides an important confirmation of the
applicability of leading twist QCD to this process.  The magnitude of
$Q^2 F_{\gamma \pi}(Q^2)$ is remarkably consistent with the predicted
form, assuming the asymptotic distribution amplitude and including the
LO QCD radiative correction with $\alpha_V(e^{-3/2} Q)/\pi \simeq
0.12$.  One
could allow for some broadening of the distribution amplitude with a
corresponding increase in the value of $\alpha_V$ at small scales.
Radyushkin \cite{Radyushkin}, Ong \cite{Ong} and Kroll \cite{Kroll}
have also noted that the scaling and normalization of the
photon-to-pion transition form factor tends to favor the asymptotic
form for the pion distribution amplitude and rules out broader
distributions such as the two-humped form suggested by QCD sum rules
\cite{CZ}.

The two-photon annihilation process $\gamma^* \gamma \to $
hadrons, which is measurable in single-tagged $e^+ e^- \to e^+ e^- {\rm
hadrons}$ events, provides a semi-local probe of
$C=+$ hadron systems $\pi^0, \eta^0, \eta^\prime, \eta_c, \pi^+ \pi^-$,
etc.  The $\gamma^* \gamma
\to \pi^+
\pi^-$ hadron pair process is related to virtual Compton
scattering on a pion target by crossing.  The leading twist amplitude is
sensitive to the
$1/x - 1/(1-x)$ moment of the two-pion distribution amplitude coupled
to two valence quarks\cite{Muller:1994fv,Diehl:2000uv}.

\section {Exclusive Two-Photon Annihilation into Hadron Pairs}

\begin{figure}[htbp]
\begin{center}
\leavevmode
{\epsfxsize=5.5in\epsfbox{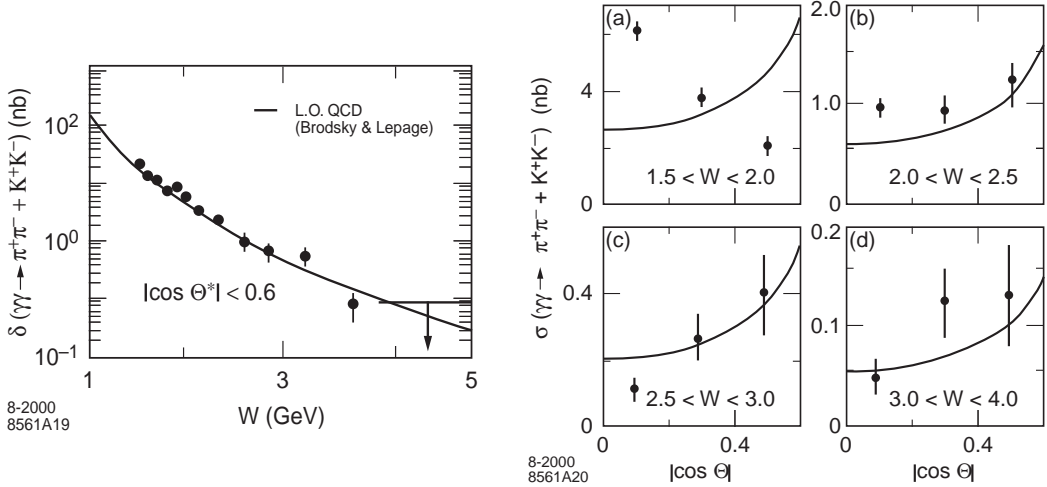}}
\end{center}
\caption[*]{Comparison of the sum of $\gamma \gamma \rightarrow \pi^+
\pi^-$ and
$\gamma \gamma \rightarrow K^+ K^-$ meson pair production cross
sections with the scaling and angular distribution of the perturbative QCD
prediction\cite{Brodsky:1981rp}.  The data are from the CLEO
collaboration\cite{Paar}.}
\label{Fig:CLEO}
\end{figure}

Two-photon reactions, $\gamma \gamma \to H \bar H$ at large s = $(k_1 +
k_2)^2$ and fixed $\theta_{\rm cm}$,
provide a particularly important laboratory for testing QCD since
these cross-channel ``Compton" processes are the simplest
calculable large-angle exclusive hadronic scattering reactions.
The helicity structure, and often even the absolute normalization can be
rigorously computed for each two-photon channel\cite{Brodsky:1981rp}.
In the case of meson pairs, dimensional counting predicts that for large
$s$, $s^4 d\sigma/dt(\gamma
\gamma \to M \bar M$ scales at fixed $t/s$ or $\theta_{\rm c.m.}$ up to
factors of $\ln s/\Lambda^2$.
The angular dependence of the $\gamma \gamma \to H \bar
H$ amplitudes can be used to determine the shape of the
process-independent distribution amplitudes, $\phi_H(x,Q)$.
An important feature of the $\gamma \gamma \to M \bar M$
amplitude for meson pairs is that the contributions of Landshoff pitch
singularities are power-law suppressed at the Born level -- even before
taking into account Sudakov form factor suppression.  There are also no
anomalous contributions from the $x \to 1$ endpoint integration region.
Thus, as in the calculation of the meson form factors, each fixed-angle
helicity amplitude can be written to leading order in $1/Q$ in the
factorized form $[Q^2 = p_T^2 = tu/s; \tilde Q_x = \min(xQ,(l-x)Q)]$:
\begin{equation}{\cal M}_{\gamma \gamma\to M \bar M}
= \int^1_0 dx \int^1_0 dy
\phi_{\bar M}(y,\tilde Q_y) T_H(x,y,s,\theta_{\rm c.m.}
\phi_{M}(x,\tilde Q_x) , \end{equation}
where $T_H$ is the hard-scattering amplitude $\gamma \gamma \to (q \bar
q) (q \bar q)$ for the production of the valence quarks collinear with
each meson, and
$\phi_M(x,\tilde Q)$ is the amplitude for
finding the valence $q$ and $\bar q$ with light-cone fractions of the
meson's momentum, integrated over transverse momenta $k_\perp < \tilde
Q.$ The contribution of non-valence Fock states are power-law suppressed.
Furthermore, the helicity-selection rules\cite{Brodsky:1981kj}
of perturbative QCD predict that
vector mesons are produced with opposite helicities to leading order in
$1/Q$ and all orders in $\alpha_s$.
The dependence in $x$ and $y$ of
several terms in $T_{\lambda, \lambda'}$ is quite similar to that
appearing in the meson's electromagnetic form factor.
Thus much of the
dependence on
$\phi_M(x,Q)$ can be eliminated by expressing it in terms of the
meson form factor.
In fact, the ratio of the
$\gamma
\gamma
\to \pi^+
\pi^-$ and $e^+ e^- \to \mu^+ \mu^-$ amplitudes at large $s$ and fixed
$\theta_{CM}$ is nearly insensitive to the
running coupling and the shape of the pion distribution amplitude:
\begin{equation}{{d\sigma \over dt }(\gamma \gamma \to \pi^+ \pi^-)
\over {d\sigma \over dt }(\gamma \gamma \to \mu^+ \mu^-)}
\sim {4 \vert F_\pi(s) \vert^2 \over 1 - \cos^2 \theta_{\rm c.m.} }
.\end{equation}
The
comparison of the PQCD prediction for the sum of $\pi^+ \pi^-$ plus $K^+
K^-$ channels with recent CLEO data\cite{Paar} is shown in Fig.
\ref{Fig:CLEO}.
The CLEO data
for charged pion and kaon pairs show a clear transition to the
scaling and angular distribution predicted by PQCD\cite{Brodsky:1981rp}
for
$W = \sqrt(s_{\gamma \gamma} > 2$ GeV.   It is clearly
important to measure the magnitude and angular dependence of the
two-photon production of neutral pions and
$\rho^+ \rho^-$ cross sections in view of the strong
sensitivity of these channels to the shape of meson distribution
amplitudes.
QCD also predicts that the production cross section for charged
$\rho$-pairs (with any helicity) is much larger that for that of
neutral $\rho$ pairs, particularly at large $\theta_{\rm c.m.}$ angles.
Similar predictions are possible for other helicity-zero mesons.

\vspace{.5cm}

Baryon pair production in two-photon annihilation is also an important
testing ground for QCD.  The only available data is the cross channel
reaction,
$\gamma p \to \gamma p$.  The calculation of $T_H$ for Compton
scattering requires the evaluation of 368 helicity-conserving tree
diagrams which contribute to
$\gamma (qqq)
\to
\gamma^\prime (qqq)^\prime$ at the Born level and a careful integration
over singular intermediate energy denominators
\cite{Farrar:1990qj,Kronfeld:1991kp,Guichon:1998xv}.
Brooks and Dixon\cite{Brooks:2000nb} have recently completed
a recalculation of the Compton process at leading order in PQCD, extending
and correcting earlier work.  It is useful
to consider the ratio $
{s^6 d\sigma/dt(\gamma p \to \gamma p)/ t^4 F^2_1 (e p \to e p)}$
where $F_1(t)$ is the elastic
helicity-conserving Dirac form factor since the power-law fall-off, the
normalization of the valence wavefunctions, and much of the
uncertainty from the scale of the QCD coupling
cancel.
The scaling and angular dependence of this
ratio
is sensitive to the shape of the proton distribution amplitudes and
appears to be consistent with the distribution amplitudes motivated by QCD
sum rules.  The normalization of the ratio at leading order is not
predicted correctly by perturbative QCD.  However, it is conceivable that
the QCD loop corrections to the hard scattering
amplitude
are
significantly larger than those of the elastic form factors in view of the
much greater number of Feynman diagrams contributing to the
Compton amplitude relative to the proton form factor.
The perturbative QCD predictions for the phase of
the Compton amplitude phase can be tested in virtual Compton scattering
by interference with Bethe-Heitler processes\cite{Brodsky:1972vv}.

A debate has
continued\cite{Isgur:1989iw,Radyushkin:1998rt,Bolz:1996sw,Vogt:2000bz}
on whether processes such as the pion and
proton form factors and elastic Compton scattering $\gamma p \to \gamma
p$ might be dominated by higher-twist mechanisms until very large momentum
transfer.  If one
assumes that the light-cone wavefunction of the pion has the form
$\psi_{\rm soft}(x,k_\perp) = A \exp (-b {k_\perp^2\over x(1-x)})$,
then the
Feynman endpoint contribution to the overlap integral at small $k_\perp$ and
$x \simeq 1$ will dominate the form factor compared to the hard-scattering
contribution until very large $Q^2$.  However, this ansatz for
$\psi_{\rm soft}(x,k_\perp)$ has no suppression at $k_\perp =0$
for any $x$; \ie, the
wavefunction in the hadron rest frame does not fall-off at all
for $k_\perp
= 0$ and $k_z \to - \infty$.  Thus such wavefunctions do not represent well
soft QCD contributions.  Endpoint
contributions are also suppressed by the QCD Sudakov form factor,
reflecting the fact that a near-on-shell quark must radiate if it absorbs
large momentum.
One can show \cite{Lepage:1980fj} that the leading
power dependence of the two-particle light-cone Fock wavefunction in the
endpoint region is
$1-x$, giving a meson structure function which falls as $(1-x)^2$ and
thus by duality a non-leading contribution to the meson form factor
$F(Q^2)
\propto 1/Q^3$.  Thus the dominant contribution to meson form factors
comes from the hard-scattering regime.
Radyushkin
\cite{Radyushkin:1998rt} has argued that the Compton amplitude is
dominated by soft
end-point contributions of the proton wavefunctions where the two photons
both interact on a quark line carrying nearly all of the
proton's momentum.  This description appears to agree with the Compton
data at least at forward angles where
$-t < 10$ GeV$^2$.  From this viewpoint, the dominance of the factorizable
PQCD leading twist contributions requires momentum transfers much
higher than those currently available.  However, the
endpoint model cannot explain the empirical success of the perturbative
QCD scaling $s^7 d\sigma/dt(\gamma p \to \pi^+ n) \sim {\rm const} $
at relatively low momentum transfer in pion photoproduction
\cite{Arr}.

\section{Conclusions}

The leading-twist QCD predictions for exclusive two-photon processes such
as the photon-to-pion transition form factor and $\gamma \gamma \to $
hadron pairs are based on rigorous factorization theorems.  The recent
data from the CLEO collaboration on $F_{\gamma \pi}(Q^2)$ and the sum of
$\gamma \gamma \to \pi^+ \pi^-$ and $\gamma \gamma \to K^+ K^-$
channels are in excellent agreement with the QCD predictions.  It is
particularly compelling to see a transition in angular dependence between
the low energy chiral and PQCD regimes.  The success of leading-twist
perturbative QCD scaling
for exclusive processes at presently experimentally accessible momentum
transfer can be understood if the effective coupling
$\alpha_V(Q^*)$ is approximately constant at the relatively
small scales $Q^*$ relevant to the hard scattering
amplitudes\cite{Brodsky:1998dh}. The evolution of the quark distribution
amplitudes in the low-$Q^*$ domain at also needs to be minimal.  Sudakov
suppression of the endpoint contributions is also strengthened if the
coupling is frozen because of the exponentiation of a double logarithmic
series.

One of the formidable challenges in QCD is the calculation of
non-perturbative wavefunctions of hadrons from first principles.  The
recent calculation of the pion distribution amplitude by
Dalley\cite{Dalley:2000dh} using light-cone and transverse lattice
methods is particularly encouraging.  The predicted form of
$\phi_\pi(x,Q)$ is somewhat broader than but not inconsistent with the
asymptotic form favored by the measured normalization of $Q^2 F_{\gamma
\pi^0}(Q^2)$ and the pion wavefunction inferred from diffractive di-jet
production.

Clearly much more experimental input on hadron wavefunctions is needed,
particularly from measurements of two-photon exclusive reactions into
meson and baryon pairs at the high luminosity
$B$ factories.  For example, the ratio 
$${{d\sigma \over dt
}(\gamma
\gamma \to \pi^0
\pi^0)
/ {d\sigma \over dt}(\gamma \gamma \to \pi^+ \pi^-)}$$
is particularly sensitive to the shape of pion distribution amplitude.
Baryon pair production in two-photon reactions at threshold may reveal
physics associated with the soliton structure of baryons in
QCD\cite{Sommermann:1992yh}. In addition, fixed target experiments can
provide much more information on fundamental QCD processes such as deeply
virtual Compton scattering and large angle Compton scattering.

\section*{Acknowledgments}

Work supported by the Department of Energy
under contract number DE-AC03-76SF00515.  I am grateful to Markus Diehl
and Hans Paar for their input to this talk and their help as session
organizers at Photon 2000.

\end{document}